\documentclass[a4paper, 12pt, oneside]{article}
\usepackage[cp1251]{inputenc}
\usepackage[english]{babel}
\usepackage{graphics, amsfonts, amsmath, bm, amssymb, array}
\usepackage[dvips]{graphicx}
\usepackage[flushleft,small]{caption2}

\newcommand{\diag}{\rm \diag\, }

\newcommand{\const}{\mathop{\rm const\, }}

\makeatother {

\begin{document}
\thispagestyle{empty} \large

\renewcommand{\refname}{\begin{center}{\bf REFERENCES}\end{center}}
\begin{center}
{\bf Constructing a Model Transport Equation for a Massless
Bose  Gas and its Analytic Solution}
\end{center}


\begin{center}
  \bf  A. V. Latyshev\footnote{$avlatyshev@mail.ru$} and
  A. A. Yushkanov\footnote{$yushkanov@inbox.ru$}
\end{center}\medskip

\begin{center}
{\it Faculty of Physics and Mathematics,\\ Moscow State Regional
University,  105005,\\ Moscow, Radio st., 10--A}
\end{center}\medskip

\begin{abstract}
A  model  kinetic  equation  is  constructed  for  the  transport of
a  massless  Bose  gas.  This  equation  is applied  to solution of
the boundary value problem  for  the transport  of radiation  in the
half-space occupied  by  a dispersive  medium  that  is in local
thermal equilibrium with  the  radiation.  It  is  shown  that the
difference in  temperature  between  the  dispersive  medium and the
incident  radiation  depends substantially  on  the character of the
scattering properties  of  the particles of the  medium.

\medskip

\medskip

Thermodynamics.
\end{abstract}

\begin{center}\bf
Introduction.  Statement  of the problem  and  basic  equations
\end{center}

In  recent  decades,  transport  theory  has  been  substantially
developed  for  Boltzmann  gas  [1,  2],  Boltzmann  plasma [3,  4],
and  Fermi  gas  [3-6].  Progress  in describing radiation transport
in  dispersive  ("turbid") media  should  also be mentioned  [7, 8].
Both  approximate and  exact  analytic methods (for different
collision -- integral  models)  have  been  used  in these
investigations.

In  contrast  to  the  case  of Boltzmann  and  Fermi  gases,
no  general  statement  of  the  transport  problem
is  known  for  Bose  gas,  although  some  particular  results
have  been  obtained  in  this  case  (e.g.,  see  [7,  8]).
At  the  same  time,  transport  phenomena  are  frequently
encountered  in  practice,  e.g.,  heat  transport  by
phonons  in  a  solid  and  radiative  transport  by  photons
in  dispersive  media.  All  of  these  processes  have  a
number  of common  properties.  In  what  follows,  we  confine
ourselves  to  the  case  of massless  Bose  systems
for which  the  chemical  potential  vanishes  in  the
thermodynamic  equilibrium.

Consider  the  problem  of constructing  a  transport equation
describing  the  evolution  of Bose  gas.  The collision  integrals
used  in  transport  theory  take into  account  either  the elastic
scattering  of photons (electromagnetic  waves)  or  their
absorption  by a  dispersive medium.  However,  if the  dispersive
medium  in which radiative  transport  takes  place  is  in  local
equilibrium with  the  photon  gas,  then  it  not  only  absorbs
but also  re-radiates  the  photons.  As  an  example,  we mention
the  case  of  radiative  transport  in  gaseous  dust clouds  in
the  cosmos,  where  precisely  this  type of radiative  transport
from  central  regions  to  the  outside  of the  cloud  results  in
cooling  and  compression  of the  cloud,  which,  by--turn,  gives
rise  to  the  star-formation process  [9].

As  an  example  of  the  construction  of a  model transport
equation,  we  consider  the  Bhatnagar --- Gross --- Krook  (BGK)
model  for  the  Boltzmann  equation. The  Boltzmann  equation  for
the  distribution function  $f(\mathbf{r},\mathbf{v}, t)$ of  the
velocity of gas molecules has the form
$$
\dfrac{\partial f}{\partial t}+\mathbf{v}\nabla f=J[f],
\eqno{(1)}
$$
where  $J[f]$  is  the  collision  integral.

In  constructing  the  model,  the  left-hand  side  of  Eq.  (1)
remains  unchanged,  whereas  the  collision integral  is
transformed  according  to  the  following  scheme. It  is  assumed
that  the  molecules  come to the equilibrium  distribution  after a
single  collision,  i.e., the  molecule  dictistribution  function
after  the  collision coincides  with  the  equilibrium distribution
function
$$
f_0=n_0\Big(\dfrac{m}{2kT_0}\Big)^{3/2}\exp\Big[-\dfrac{m}{2kT_0}
(\mathbf{v}-\mathbf{u})^2\Big],
$$
where $n_0,T_0$ and $\mathbf{u}$ are defined below.

In this case, the Boltzmann equation becomes
$$
\dfrac{\partial f}{\partial t}+\mathbf{v}\nabla f=\nu(v)(f_0-f),
\eqno{(2)}
$$
where $\nu(v)$ is  the  effective  collision  frequency.

The  Boltzmann  equation  (1)  is  compatible  with  the
conservation  laws  for  the  number  of molecules,
momentum,  and  energy.  The  model  equation  (2)  must
possess  similar  properties.  This  requirement  leads
to  the  uniquely  determined  parameters  $n_0,T_0$ and $\mathbf{u}$
(for  more  detail,  see  [10]).  In  the  case  $\nu(v)  =  \const$,
the  parameters $n_0,T_0$ and $\mathbf{u}$ coincide  with  the
average  molecule  concentration,  temperature,  and  velocity,
respectively.

We consider  the  case  of a  massless  Bose  gas,
meaning  a  photon  gas.  The  case  of a  phonon  gas  differs  only
in  some  details,  which  are  indicated  in  the
course  of the  presentation.

The  state  of a  photon  gas  is described  by  the  distribution
function  $f_i(\omega,\mathbf{r},\mathbf{n})$,  where $\omega$
is  the  frequency, $\mathbf{n}$
is  the  direction  of photon  propagation
($\mathbf{n}  = \dfrac{c}{\omega}\mathbf{k}$, $\mathbf{k}$
is  the  wave  vector  and  $c$  is  the  velocity  of  light),  and
the  subscript  $i$  ($i  =  1,2$)  indexes  the  polarization  direction.
For  a  phonon  gas,  there  are  three  possible
types of polarization  and,  therefore,  the  subscript  $i$
assumes  the  values  $i  =  1, 2, 3$  in  this  case.  The  transport
equation  for  the  distribution  function
$f_i(\omega, \mathbf{r},\mathbf{n})$  has  the  form
$$
\dfrac{\partial f}{\partial t}+c\mathbf{n}\nabla f=J[f].
\eqno{(3)}
$$

Here  the  functional  $J[f]$  describes  the  absorption and
re-radiation  process  for  phonons  in  the  dispersive medium
where  they  propagate.  We note  that  in  contrast  to  photons,
the phonons  interact  with  one  another. In  this  case,  $J[f]$
means  the  collision  integral  for  phonons. In  Eq.  (3),  the
parameter  $c$  is  the  velocity  of light  for  the  photon  gas
and  the  velocity  of sound  in  the case  of  the  phonon  gas (on
the  condition  that  the propagation  velocities  of  the
longitudinal  and transverse oscillation  coincide).

Let  a  photon  gas  be  in  local  equilibrium  with  a
dispersive  medium.  In  this  case,  the  distribution
function  of photons  re-radiated  by  the  medium  coincides
with  the  Planck  function
$$
f_P=\Bigg[\exp \dfrac{\hbar\omega}{kT}-1\Bigg]^{-1},
$$
where  $T$  is  a  parameter  defined  below.
Equation  (3)  takes  the  form
$$
\dfrac{\partial f_i}{\partial t}+c\mathbf{n}\nabla f_i=
\nu(\omega)(f_P-f_i),
\eqno{(4)}
$$
where  $\nu(\omega)$  is  the  effective  froquency  of  the
phonon  collisions.  For photons,  this  quantity  is  related  to
the effective  absorption  cross-section for  the dispersing particles,
namely,
$$
\nu(\omega)=c\int n(\mathbf{r})\sigma_r(\omega)dr,
$$
where  $n(\mathbf{r})$  is  the  radial  distribution  of  the
particles  (on  the  condition  that  they  have a  spherical  form)  and
$\sigma_r(\omega)$  is  the  absorption  cross-section  for
a  particle  of  radius  $r$.

Note  that  in  the  general  case,  it  is necessary  to  distinguish
between  photon  scattering  and  re-radiation.
For  simplicity,  we do  not  discuss  this  question
in  the  present  paper.

In  general,  the  number  of  photons  and  their  momenta  are  not
preserved  under  re-radiation.  At  the
same  time,  the  photon  gas energy must  be  invariant  in  local
thermodynamic  equilibrium  between  the  photon
gas  and  the  dispersion  system.
This  means  that  the  integral  on  the  right-hand  side
of Eq.  (4)  vanishes,
$$
\int \sum\limits_{i=1}^{2}\omega^3\nu(\omega)(f_P-f_i)d\mathbf{n}
d\omega=0.
\eqno{(5)}
$$

Relationship  (5)  is  the one whichn  determines  the  parameter
$T$.

We consider  the  case  where  the  Bose  system  is  in  a
nearly  equilibrium  state.  Here  the  distribution  function  $f_i$
can  be  represented  in  the  form
$$
f_i=f_P^{(0)}+\psi_i,
\eqno{(6)}
$$
where  $\psi_i$  is  a  linear  correction  to  the  Planck
distribution  $f_P^{(0)}$.

In  what  follows,  we  assume  that  the  radiation  is  not
polarized  and  the  distribution  function  does  not
depend  on  $i$.  Therefore,  we  omit  the  subscript  $i$  in
the  distribution  functions  $f_i$  and  $\psi_i$.
The  closeness  to
equilibrium  means  that  $T_0$  and  the  parameter  $T$,
which  has  the  meaning  of  the  effective  temperature,  vary
weakly  in  the  gas  volume,  which  is  equivalent  to  the  relationship
$$
|\delta|\ll 1,
\eqno{(7)}
$$
where  $\delta=\dfrac{T-T_0}{T_0}$, and $T_0$  is  the  temperature
of  the  Bose  gas  at  some  point  in  the  volume.

Under  the condition  (7),  we  can  linearize  the  Planck
function $f_P(\delta)$  with  respect  to  the  parameter $\delta$,
$f_P(\delta)=f_P(0)+f'_P(0)\delta$, where
$$
f_P(\delta)=\Bigg[\exp \Big(\dfrac{\hbar \omega}{k(T_0+\delta T_0)}\Big)
-1\Bigg]^{-1},
$$
or, in explicit form,
$$
f_P=f_P^{(0)}+\dfrac{\hbar \omega}{kT_0^2}E\Big(\dfrac{\hbar\omega}
{kT_0}\Big)(T-T_0).
\eqno{(8)}
$$

Here $E(x)$ is the Einstein function,
$$
E(x)=\dfrac{e^x}{(e^x-1)^2}.
$$

On  substituting  (6)-(8)  into  Eq.  (4),  we  derive
$$
\dfrac{\partial \psi}{\partial t}+c\mathbf{n}\nabla \psi=
\nu(\omega)\Bigg[\dfrac{\hbar \omega}{kT_0^2}E\Big(\dfrac{\hbar\omega}
{kT_0}\Big)(T-T_0)-\psi(\omega,\mathbf{r},\mathbf{n})\Bigg].
\eqno{(9)}
$$

Equation  (9)  is  the  linearized  form  of the transport  equation
(4).

It  follows  from  (5)  that  the  difference  $T-To$  is  given  by
the  formula
$$
T-T_0=\dfrac{kT_0^2}{\hbar}\Bigg[\int \nu(\omega)\omega^4
E\Big(\dfrac{\hbar \omega}{kT_0}\Big)d\omega d\mathbf{n}\Bigg]^{-1}
\int \nu(\omega)\omega^3\psi d\omega d\mathbf{n}.
$$

Equation  (9)  shows  that  it  is  convenient  to  introduce  a
new  function  $\varphi$  instead  of $\psi$,
$$
\psi(\omega,\mathbf{r},\mathbf{n})=
\omega E\Big(\dfrac{\hbar \omega}{kT_0}\Big)
\varphi(\omega,\mathbf{r},\mathbf{n}),
$$
and, then,  Eq.  (9)  can  be  rewritten as
$$
\dfrac{\partial \varphi}{\partial t}+c\mathbf{n}\nabla \varphi=$$$$=
\dfrac{\nu(\omega)}{l}\int\nu(\omega')
{\omega'}^4E\Big(\dfrac{\hbar \omega'}{kT_0}\Big)
\varphi(\omega',\mathbf{r},\mathbf{n})\, d\omega' d\mathbf{n}-
\nu(\omega)\varphi(\omega,\mathbf{r},\mathbf{n}),
\eqno{(10)}
$$
where
$$
l=\int\nu(\omega')
{\omega'}^4E\Big(\dfrac{\hbar \omega'}{kT_0}\Big)\,d\omega'd\mathbf{n}.
$$

We consider  the  case  of a  stationary process  and  assume  that
the  function  $\nu(\omega)$  can  be  approximated  by a  power
function,  $\nu(\omega)=\nu_0\omega^\alpha$. Note  that  the  case
$\alpha  =  2$ corresponds  to  the  absorption of electromagnetic
radiation by particles whose size  is small  in comparison with the
wavelength.  For  large-sized particles, $ \alpha  =  0$. We
introduce  the  dimensionless  variables
$$
\omega^*=\dfrac{\hbar \omega}{kT_0} \quad \text{and}\quad
\mathbf{r}^*=\dfrac{\nu_0}{c}\Big(\dfrac{kT_0}{\hbar}\Big)^\alpha
\mathbf{r}.
$$

In  this  case,  Eq.  (10)  takes  the  form
(here  and  henceforth,  the  asterisks  in  the  variables  are  omitted)
$$
\mathbf{n}\nabla \varphi=\dfrac{\omega^\alpha}{4\pi l_0(\alpha)}\int
{\omega'}^{\alpha+4}E(\omega')\varphi(\omega',\mathbf{r},\mathbf{n})\,
d\omega' d\mathbf{n}-\omega^\alpha \varphi(\omega,\mathbf{r},\mathbf{n}),
\eqno{(11)}
$$
where
$$
4\pi l_0(\alpha)=
\int \omega^{\alpha+4}E(\omega)d\omega d\mathbf{n}=
4\pi\int\limits_{0}^{\infty} \omega^{\alpha+4}E(\omega)d\omega,
$$
hence
$$
l_0(\alpha)=\int\limits_{0}^{\infty}
\omega^{\alpha+4}E(\omega)d\omega.
$$

We note  that  it is assumed  in  the  above  calculations that
$\nu(\omega)$ does  not depend  on  the  coordinate,  though,
generally,  this  is  not  the  case.  This  fact  can  be  taken
into  account  by  introducing  the  optical  length  [8]. We  do
not  dwell on  this  in what  follows.

We consider  the  one-dimensional problem  in which
$\varphi=\varphi(x,v,\omega)$,  where  $v=n_x$v,  i.e.,
$v$  is  the  cosine of
the  angle  between  the  direction  of  the  quantum motion  and
the  $x$-axis.  Here,  Eq.  (11)  can  be  rewritten  as
$$
v\dfrac{\partial }{\partial x}\varphi(x,v,\omega)+\omega^\alpha
\varphi(x,v,\omega)=
$$
$$
=\dfrac{\omega^\alpha }{2l_0(\alpha)}
\int\limits_{0}^{\infty}
{\omega'}^{\alpha+4}E(\omega')d\omega'\int\limits_{-1}^{1}
\varphi(x,v',\omega')dv',
\eqno{(12)}
$$
where
$$
l_0\equiv l_0(\alpha)=
\int\limits_{0}^{\infty}\omega^{\alpha+4}E(\omega)d\omega=
\int\limits_{0}^{\infty}\dfrac{e^\omega \omega^{\alpha+4}}
{(e^\omega-1)^2}d\omega.
$$

In  the  transition  to  the Eq.  (12),  we  integrated  with
respect to  the  azimuthal  angle  from  $0$  to  $2\pi$  because
the function $\varphi$  does  not  depend  on  this  angle.

We consider  a  concrete  Milne-type  problem.  Let  a  dispersive
medium  occupy  the  half-space  $x  >  0$  and
let  there  be  a  temperature  gradient  along  the  $x$-axis
in  the  medium  far  from  the  boundary. We assume  that
radiation at  equilibrium with  the  temperature $T_0$ comes
from  the  half-space $x  <  0$  into  the medium  and  the
total  radiation outgoing from  the medium  is absorbed  somewhere
and  does  not  come back.  The  temperature
far  from  the  boundary  has  the  form
$$
T=T_1+Kx,
\eqno{(13)}
$$
where  $K$  is  the  temperature  gradient.  Our  aim  is  to  find
the  relationship  between  $T_0$,  $T_1$,  and  $K$.  This  can
be  written  in  the  linear  approximation  as
$$
T_1-T_0=fK,
\eqno{(14)}
$$
where  $f$  is  an  unknown  coefficient depending  on  the
properties  of the  medium.

In  application  to  the  cosmic  gaseous  dust  cloud,  this
statement  of  the  problem  means  the  following.
The  radiation  at  equilibrium with  the  temperature  $T_0$
close  to  $3 \rm K$  is  incident on  the  boundary  of a  gaseous
dust  cloud.  Under  cooling  of  the  cloud,  energy  is  transformed
and  radiated  outward.  Therefore,  there  is  a
temperature  gradient  in  the  cloud.  We  need  to  find
the  relationship  between  this  gradient,  which  is defined
by  the  heat  flow,  the  temperature  $T_0$,  and  the  temperature
at  the  boundary.  This  relationship  is  given  by
formula  (14)  with  an  unknown  coefficient  $f$  that  must
be  determined.

It  is  easy  to  show  that  the  following  two  discrete
modes  are  solutions  to  Eq.  (12)  (see  Sec.  1  below):
$$
\varphi_+(x,v,\omega)=1,\qquad
\varphi_-(x,v,\omega)=x-\dfrac{v}{\omega^\alpha}.
$$

The  boundary  conditions  are  as  follows:
$$
\varphi(0,v,\omega)=0,\qquad 0<v<1,
\eqno{(15)}
$$
$$
\varphi(x,v,\omega)=K_0+K\Big(x-\dfrac{v}{\omega^\alpha}\Big)+o(1),\quad
x\to+\infty, \quad -1<v<0,
\eqno{(16)}
$$
where  $K$  is  the  temperature  gradient  (see  (13))  and
$K_0  =  fKT_0^{-1}$  is  an  unknown  coefficient  proportional
to  the  relative  temperature  gradient.
Note  that  a  similar  question  for  the  Boltzmann  gas  can
be  stated  as
the  Smolukhowski problem  on  the  discontinuity  of temperature.

\begin{center}
  \bf 1. Eigenfunctions  corresponding
  to  continuous  and  discrete  spectra
\end{center}

It  can  be  clearly  seen  from  Eq.  (12)  and  boundary
conditions  (15),  (16)  that  the  desired  function  $\varphi$
depends  on  two  variables,  $x$  and  $\mu  =  v\omega^{-\alpha}$.
Let  us  perform  the  change  of variables  $v  = \mu{\omega'}^\alpha$
and  $\omega=\omega'$ in  Eq.  (12).  The  corresponding
Jacobian  is  equal  to  ${\omega'}^\alpha$.
Consequently,  Eq.  (12)  can  be  rewritten  in  the
new  variables  as
$$
\mu\dfrac{\partial \varphi}{\partial x}
+\varphi(x,\mu)=
\dfrac{1}{2l_0(\alpha)}\int\limits_{0}^{\infty}\omega^{2\alpha+4}
E(\omega)d\omega\int\limits_{-1/\omega^{\alpha}}^{1/\omega^{\alpha}}
\varphi(x,\mu')d\mu'.
\eqno{(1.1)}
$$

Accordingly,  under  the  above  transformation,
the  boundary  conditions  (15),  (16)  for Eq.  (1.1)  take  the  form
$$
\varphi(0,\mu)=0, \qquad 0<\mu<\infty,
\eqno{(1.2)}
$$
$$
\varphi(x,\mu)=K_0+K(x-\mu)+o(1),\quad x\to+\infty,\quad
-\infty<\mu<0.
\eqno{(1.3)}
$$

On  replacing $\mu'$  by $\omega^{-\alpha}\mu'$  in  Eq.  (1.1),
we  can  rewrite  it  as
$$
\mu\dfrac{\partial}{\partial x}\varphi(x,\mu)+\varphi(x,\mu)=$$$$=
\dfrac{1}{2l_0(\alpha)}\int\limits_{0}^{\infty}\omega^{\alpha+4}
E(\omega)d\omega\int\limits_{-1}^{1}\varphi(x,\omega^{-\alpha}\mu')
d\mu'.
\eqno{(1.4)}
$$

Let  us  perform  the  separation  of variables
in  Eq.  (1.4)  as  follows:
$$
\varphi_\eta(x,\mu)=\exp(-\dfrac{x}{\eta})\Phi(\eta,\mu),\qquad
\eta\in \mathbb{C}.
$$

As  a  result,  we  derive  the  characteristic  equation
$$
(\eta-\mu)\Phi(\eta,\mu)=\dfrac{1}{2l_0(\alpha)}\eta
\int\limits_{0}^{\infty}\omega^{\alpha+4}
E(\omega)d\omega\int\limits_{-1}^{1}\Phi(\eta,\omega^{-\alpha}\mu')
d\mu'.
\eqno{(1.5)}
$$

We denote
$$
n(\eta)=\int\limits_{0}^{\infty}\omega^{\alpha+4}
E(\omega)d\omega\int\limits_{-1}^{1}\Phi(\eta,\omega^{-\alpha}\mu')
d\mu'
\eqno{(1.6)}
$$
and rewrite Eq. (1.5) in the form
$$
(\eta-\mu)\Phi(\eta,\mu)=\dfrac{1}{2l_0(\alpha)}\eta n(\eta).
\eqno{(1.7)}
$$

Equation  (1.7)  implies  that  the  eigenfunctions  corresponding
to  the  continuous  spectrum  are  the  distributions  [11]
$$
\Phi(\eta,\mu)=\dfrac{1}{2l_0(\alpha)}\eta
n(\eta)P\dfrac{1}{\eta-\mu}+g(\eta)\delta(\eta-\mu),\quad
\eta,\mu\in (-\infty,+\infty).
\eqno{(1.8)}
$$

Here  $P$  symbolizes  the  principal  value  of  the  integral,
$\delta(x)$  is  the  Dirac  delta  function,  and  $g(\eta)$  can  be
found  from  the  normalization  condition  for  (1.6).
Substituting  (1.8)  into  (1.6),  we  obtain
$$
n(\eta)\lambda(\eta)=g(\eta)\xi_\alpha(\eta),
\eqno{(1.9)}
$$
where
$$
\xi_\alpha(\eta)=\int\limits_{0}^{1/\eta^{\alpha}}
\omega^{2\alpha+4}E(\omega)d\omega,
$$
and the dispersion function
$$
\lambda(z)=1+\dfrac{z}{2l_0(\alpha)}\int\limits_{0}^{\infty}
\omega^{2\alpha+4}E(\omega)d\omega \int\limits_{-1}^{1}\dfrac{d\mu}
{\mu-\omega^\alpha z}
$$
of  the  problem  can  be  expressed  via  the  Case
dispersion  function  [12]
$$
\lambda_C(z)=1+\dfrac{1}{2}z\int\limits_{-1}^{1}\dfrac{d\mu}{\mu-
\omega^\alpha z}
$$
in the following way
$$
\lambda(z)=\dfrac{1}{l_0(\alpha)}\int\limits_{0}^{\infty}\omega^{2\alpha+4}
E(\omega)\lambda_C(\omega^\alpha z)d\omega.
\eqno{(1.10)}
$$

The  boundary values of the  function $\lambda_C(\omega^\alpha z)$
on  the  edges of the  slit
$$
\Delta_\alpha=\Big(-\dfrac{1}{\omega^{\alpha}},
\quad\dfrac{1}{\omega^{-\alpha}}\Big)
$$
are determined by  the  Sokhotski  formulas,
$$
\lambda_C^{\pm}(\omega^\alpha\mu)=\left\{ \begin{array}{cc}
\lambda_C(\omega^\alpha\mu)\pm
i\dfrac{\pi}{2}\omega^\alpha\mu, & \mu\in \Delta_\alpha, \\
\lambda_C(\omega^\alpha\mu), & \mu\notin \Delta_\alpha.
\end{array}\right.
\eqno{(1.11)}
$$

In  view  of  (1.10)  and  (1.11),  the  boundary  values  of
$\lambda(z)$  are  related  by  the  formula
$$
\lambda^{\pm}(\mu)=\lambda(\mu)
\pm i \dfrac{\pi}{2}\dfrac{\xi_\alpha(\mu)}{l_0(\alpha)}, \qquad
\mu\in \mathbb{R}.
$$

With  the  help  of expression  (1.9),  we  bring  formula  (1.8)
for  the  eigenfunctions  to  the  form
$\Phi(\eta,\mu)=\tilde{\Phi}(\eta,\mu)n(\eta)$,
where
$$
\tilde{\Phi}(\eta,\mu)=\dfrac{1}{2l_0(\alpha)}\eta
P\dfrac{1}{\eta-\mu}+\dfrac{\lambda(\eta)}{\xi_\alpha(\eta)}\delta(\eta-\mu)
$$
is  the  eigenfunction  corresponding  to the  normalization
condition
$$
n(\eta)\equiv 1.
$$

It  can  be shown  that  $\lambda(z)$  has  a  second--order zero  at
$z=\infty$  with  which  two  solutions  to the  original
equation  (1.4)  are  associated,  namely,
$$
\varphi_+(x,\mu)=1\quad \text{and}\quad \varphi_-(x,\mu)=x-\mu.
$$

\begin{center}
\bf  2. Eigenfunction  expansion  for   the  solution
\end{center}

{\bf \large Theorem.}
{\sl  The  boundary  value problem  (1.2)-(1.4)  has  a  unique
solution,  whose  expansion  in  terms
of the  eigenfunctions  of  the  characteristic  equation  has  the
form}
$$
\varphi(x,\mu)=K_0+K(x-\mu)+\int\limits_{0}^{\infty}
e^{-x/\eta}\tilde{\Phi}(\eta,\mu)n(\eta)d \eta.
\eqno{(2.1)}
$$

Here  the  unknowns  are  the  coefficient $K_0$  associated with
the  discrete  spectrum  and  the  function  $n(\eta)$, which  is
the  coefficient  corresponding  to  the  continuous  spectrum. The
expansion  (2.1)  is  interpreted  in  the classical  sense, namely,
$$
\varphi(x,\mu)=K_0+K(x-\mu)+\hspace{6cm}
$$
$$
+\dfrac{1}{2l_0(\alpha)}\int\limits_{0}^{\infty}
e^{-x/\eta}\dfrac{\eta n(\eta)}{\eta-\mu}d\eta+
\dfrac{\lambda(\mu)}{\xi_\alpha(\mu)}e^{-x/\mu}\theta_+(\mu),
\eqno{(2.2)}
$$
where $\theta_+(\mu)$  is  the  characteristic  function  of
the  positive  half-line,  i.e.,  $\theta_+(\mu) = 1$  if
$\mu\in (0, +\infty)$  and $\theta_+(\mu) =0$  if $\mu\notin (0, +\infty)$.

{\bf \large Proof}.  Using boundary  condition  (1.2),  we pass  from
expansion  (2.2)  to  the  following singular  integral
equation  with  the  Cauchy  kernel  [13]:
$$
K_0-K_1\mu+\dfrac{1}{2l_0(\alpha)}\int\limits_{0}^{\infty}
\dfrac{\eta n(\eta)d\eta}{\eta-\mu}+
\dfrac{\lambda(\mu)}{\xi_\alpha(\mu)}n(\mu)=0, \quad \mu>0.
\eqno{(2.3)}
$$

Let  us  introduce  the  auxiliary  function
$$
N(z)=\int\limits_{0}^{\infty}\dfrac{\eta n(\eta)d\eta}{\eta-z}.
\eqno{(2.4)}
$$

Multiplying  both  sides  of Eq.  (2.3)  by
$$
\dfrac{2}{l_0(\alpha)}[\lambda^+(\mu)-\lambda^-(\mu)]=
2\pi i \mu \xi_\alpha(\mu)
$$
(2.4)
and  using  the  boundary  values $N^{\pm}(\mu)$  and
$\lambda^{\pm}(\mu)$,  we  reduce  Eq.  (2.3)  to  the  Riemann
boundary  problem
$$
\lambda^+(\mu)[N^+(\mu)+2l_0(\alpha)(K_0-K_1\mu)]=
$$
$$
=
\lambda^-(\mu)[N^-(\mu)+2l_0(\alpha)(K_0-K_1\mu)], \quad \mu>0.
\eqno{(2.5)}
$$

The  coefficient  in  the  boundary  problem  (2.5)  is  the  function
$$
G(\mu)=\dfrac{\lambda^-(\mu)}{\lambda^+(\mu)}.
$$
Consider  the  following
factorization  problem  for  the  coefficient  $G(\mu)$
$$
\dfrac{X^+(\mu)}{X^-(\mu)}=\dfrac{\lambda^+(\mu)}{\lambda^-(\mu)},
\qquad \mu>0.
\eqno{(2.6)}
$$

To  solve  this  problem,  we  calculate  the  index
$\varkappa$   of  the  coefficient $G(\mu)$,
$$
\varkappa=\dfrac{1}{2\pi}\Big[\arg G(\mu)\Big]_{(0,+\infty)}.
$$
where  the expression  $[\arg G(\mu)]_{(0,+\infty)}$ denotes the
increment  of  the  function   in  the  brackets  under  variation
of the argument from $0$ to $+\infty$.

Let  $\theta_\alpha(\mu)=\arg \lambda^+(\mu)$ be the  principal
value  of the argument  of $\lambda^+(\mu)$,
$$
\theta_\alpha(\mu)=\arctan\dfrac{\displaystyle2\int\limits_{0}^{\infty}
\omega^{\alpha+4}E(\omega)\lambda_C(\omega^\alpha \tau)d\tau}
{\displaystyle\pi \tau
\int\limits_{0}^{\infty}\omega^{2\alpha+4}E(\omega)d\omega}.
$$

Since  $\overline{\lambda^+(\mu)}=\lambda^-(\mu)$ (where  the  bar
symbolizes  complex  conjugation),  we  have
$G(\mu)=\exp(-2{\it _\alpha(\mu)})$, and,  therefore,
$$
\varkappa=-\dfrac{1}{\pi}[\theta_\alpha(\mu)]_{(0,+\infty)}=
-\dfrac{1}{\pi}[\theta_\alpha(+\infty)-\theta_\alpha(0)]=-1.
$$

As  a  solution  to  problem  (2.6),  we  take  the  function
$$
X(z)  =  \dfrac{1}{z} \exp V(z),
$$
which  does  not  vanish  at  zero,
where
$$
V(z)  = \dfrac{1}{\pi}\int\limits_{0}^{\infty}\dfrac{\theta_\alpha(\tau)-
\pi}{\tau-z}d\tau.
$$
We  now  return  to  the problem  (2.5)  and  apply  (2.6)  to
transform it  into  the  problem  of  determining  an analytic
function  from  the  following condition
$$
X^+(\mu)[N^+(\mu)+2l_0(\alpha)(K_0-K_1\mu)]=$$$$+
X^-(\mu)[N^-(\mu)+2l_0(\alpha)(K_0-K_1\mu)], \qquad \mu>0.
\eqno{(2.7)}
$$

Taking  into  account  the  behavior  of  the  functions  entering
(2.7),  we  derive  the  general  solution  to  this
problem  in  the  form
$$
N(z)=-2l_0(\alpha)(K_0-Kx)+\dfrac{C_0}{X(z)},
\eqno{(2.8)}
$$
where  $C_0$  is  an  arbitrary  constant.

In  order  to  use  the function  (2.8)  as  the  auxiliary function
$N(z)$  introduced by the formula  (2.4),  we  remove  the pole  of
the solution  (2.8)  at  $z  =\infty$  and  make  the  limit of
(2.8) at  the  point $z  =\infty$  equal  to  zero  by  setting
$$
C_0=-2l_0(\alpha)K, \qquad K_0=V_1(\alpha)K,
\eqno{(2.9)}
$$
$$
V_1(\alpha)=-\dfrac{1}{\pi}\int\limits_{0}^{\infty}
[\theta_\alpha(\mu)-\pi]\,d\mu.
\eqno{(2.10)}
$$

The  coefficient  $n(\eta)$  for  the  continuous  spectrum
can  be  found  from  Eq.  (2.8),  namely,
$$
2\pi i \eta n(\eta)=-2l_0(\alpha)K\Bigg(\dfrac{1}{X^+(\eta)}-
\dfrac{1}{X^-(\eta)}\Bigg).
$$

Thus,  the  coefficients  in  (2.1)  have  been  uniquely  defined.
The  uniqueness  of the expansion  (2.1)  is  proved by  the  fact
that  there  can  be  no  nontrivial  expansion of zero  with
respect  to  the  eigenfunctions  of  the characteristic  equation.
Obviously,  expansion  (2.1) automatically  satisfies to  the
boundary condition  (1.3)  and  (1.2) holds  by  construction.
Direct verification  shows  that the expansion  (2.1)  satisfies to
the Eq. (1.4). The  theorem  is proved.

\begin{center}
  \bf 3.  Numerical  results.  Discussion  and  conclusion
\end{center}

First,  we note  that  we  can  find  the  exact  value  of the
integral (2.10)  for  $\alpha  =  0$.  Indeed,  if $\alpha=0$, then
it  can be  seen from the formula  (1.10)  that  the  dispersion
function $\lambda(z)$ coincides  with  the  Case  dispersion
function $\lambda_C(z)$.  In  this  case,
$$
V_1^\circ=-\dfrac{1}{\pi}\int\limits_{0}^{\infty}
{\rm arccot}\dfrac{\pi \tau}{2\lambda_C(\tau)}d\tau=0.71045.
\eqno{(3.1)}
$$

Let  us  find  an  approximate  value  of the integral  (2.10)  for
$\alpha  >  0$.  Applying  the  saddle-point  method, we  replace
$\lambda(z)$  by  an  approximating  function $\lambda(z)$. The
expression  (1.10)  implies  that  the  saddle  point $\omega_0$ is
determined  by  the  equation
$$
e^{\omega_0}=\dfrac{\alpha+4+\omega_0}{\alpha+4-\omega_0}.
\eqno{(3.2)}
$$

The  approximate  solution $\tilde{\omega}_0$ to  Eq.  (3.2)  for
$\alpha+4>1$ can  be  calculated  by  the  formula
$$
\tilde{\omega}_0=(\alpha+4)(1-2e^{-\alpha-4}).
\eqno{(3.3)}
$$

Note  that  formula  (3.3)  provides  a  satisfactory  approximation
even  for  $a  =  0$.  Indeed,  in  this  case  $\omega_0=
3.83002$  and  $\tilde{\omega}_0$=$3.85347$.
For  $\alpha  =  2$,  we  have  $\omega_0$=$5.96941$  and
$\tilde{\omega}_0 = 5.97025$,  i.e.,  formula  (3.3)  gives
a  nearly  exact  value  of  the  saddle  point.  According  to
the  saddle-point  method,  we  have
$$
\tilde{\lambda}(z)=\lambda_C(\omega_0^\alpha z)
$$
and,  therefore,
$$
\tilde{V_1}=-\dfrac{1}{\pi}\int\limits_{0}^{1/\omega_0^{\alpha}}
[\tilde\theta_\alpha(\tau)-\pi]d\tau=-\dfrac{1}{\pi}
\int\limits_{0}^{1/\omega_0^{\alpha}}{\rm arccot}
\dfrac{\pi \omega_0^\alpha\tau}{2\lambda_C(\omega_0^\alpha \tau)},
$$
whence
$$
\tilde V_1=\omega_0^{-\alpha}V_1^\circ.
\eqno{(3.4)}
$$

Note  that  formula  (3.4)  with  $A=0$  exactly  implies the
formula (3.1)  and  for  $\alpha  =  2$, we  have  $\tilde V_1 =
0.01994$. The  decrease  in  $V_1$ with  increasing  $\alpha$
relates to  the fact  that  the  contribution  from  high
frequencies  (large values  of $\omega$)  is  reduced  during  the
transport  process in  the  dispersive  medium  due  to  the  faster
growth  of the collision  frequency $\nu(\omega)$)  with  increasing
$\alpha$. At the  same  time,  the  radiation  from  the  boundary
surface is not  limited  by  this  factor  nor  does  the
high-frequency contribution  depend  on  $\alpha$.

The  result  is  that
the  relative  outflow  of energy  from  the  surface  increases
together  with  $\alpha$  (as  compared  to  the  volumetric
outflow).  As  a  result,  the  coefficient $V_1$, which  characterizes
the  temperature  drop  between  the  boundary
of the  exterior medium  and  the  incident  radiation,  decreases.

In  conclusion,  we  note  that  in  the  present  paper,
a  generalization  of  the  Case --- Zweifel  classical  theory
[12]  to  the  transport  equation  for  a  massless  Bose  gas  was
considered  for  the  first  time.  It  turns  out
that  the  suggested method  permits  transport  equations with
a  double  integral on  the  right-hand  side  to  be
solved  analytically.

This  work  is  supported  by  the  Russian  Foundation
for Basic  Research,  Grant  No.  97-01-00333.

\end{document}